\documentclass[pra, aps, english, twocolumn, hyperref,floatfix, superscriptaddress,showpacs]{revtex4}
\usepackage{graphicx}
\usepackage{amssymb, amsmath, amsfonts, color, rotating, multirow, graphicx, bm}
\usepackage[dvipdfm,bookmarks=false,pdfstartview=FitH,hyperindex=true, colorlinks, linkcolor=blue, citecolor=blue]{hyperref}

\begin{document}
\author{Ran Wei}
\affiliation{Hefei National Laboratory for Physical Sciences at Microscale and Department
of Modern Physics, University of Science and Technology of China, Hefei, Anhui 230026, China}
\affiliation{Laboratory of Atomic and Solid State Physics, Cornell University, Ithaca, NY, 14850}
\author{Erich J. Mueller}
\affiliation{Laboratory of Atomic and Solid State Physics, Cornell University, Ithaca, NY, 14850}
\title{Majorana fermions in one-dimensional spin-orbit coupled Fermi gases}
\date{\today}
\begin{abstract}
We theoretically study trapped one-dimensional Fermi gases in the presence of spin-orbit coupling
induced by Raman lasers. The gas changes from a conventional (non-topological) superfluid
to a topological superfluid as one increases the intensity of the Raman lasers above a critical
chemical-potential dependent value.
Solving the Bogoliubov-de Gennes equations self-consistently, we calculate the density of states
in real and momentum space at finite temperatures.
We study Majorana fermions (MFs) which appear at the boundaries between topologically trivial and topologically non-trivial regions.
We linearize the trap near the location of a MF,
finding an analytic expression for the localized MF wavefunction and
the gap between the MF state and other edge states.
\end{abstract}
\maketitle

\section{introduction}
Majorana fermions (MFs), exotic excitations which are their own antiparticles, have attracted a great deal of
attention recently \cite{Mf}. Condensed matter systems with MFs possess degeneracies that
are intrinsically nonlocal, and can be manipulated geometrically.
They can, in principle, be used to make a robust quantum computer \cite{Sarma2008}.
Condensed matter theorists have proposed various ways to explore MFs during
the past several years \cite{Read2000,Kitaev2001,SSthoery,Fuliang2008,Oreg2010,braiding}.
Four experimental groups have recently reported evidence of MFs
in semiconducting wires on superconducting substrates \cite{SSexperiment}.
In those experiments, spin-orbit (SO) coupling was important.
Here we study MFs in a related cold atom system.

Two groups \cite{Zhang2012,Zwierlein2012} have successfully generated SO coupled Fermi gases based on
a Raman technique pioneered by Spielman's group at NIST \cite{Spielman2011}.
Several theoretical groups have proposals for creating
and probing MFs in these SO coupled Fermi gases \cite{CAtheory,Zoller2011,Huihu2012a,Huihu2012b}.
We build upon the studies of Jiang \textit{et al.} \cite{Zoller2011} and Liu \textit{et al.} \cite{Huihu2012a},
which find MFs in a 1D geometry.

We study a 1D (pseudo) spin-$1/2$ Fermi gas with point interactions.
In the presence of Raman lasers, the energy spectrum has two helical bands.
We study this two-band model in a harmonic trap.
Solving the Bogoliubov-de Gennes (BdG) equations self-consistently, we calculate the density of states (DOS)
in real and momentum space at finite temperatures.
We linearize the trap near the location of a MF,
finding an analytic expression for the localized MF wavefunction and
the gap between the MF state and other edge states.

Our numerical calculations extend the similar studies of Ref. \cite{Huihu2012a}.
We explore a larger range of temperatures, and delve deeper into the physics near the MFs.
We also investigate a truncated one-band model.

One concern with mean-field calculations such as ours,
is that they are unable to capture the large phase fluctuations found
in 1D. As shown by Ref. \cite{multiwires}, the MF physics is robust against these fluctuations. Moreover, an actual
experiment would be performed on a bundle of weakly coupled tubes \cite{Hulet2010}. This latter
setting also avoids issues of number conservation \cite{multiwires}.
Our 1D model faithfully describes the properties of a single tube within such a bundle
when the tunneling is weak.

This paper is organized as follows. In Sec. II, we discuss the homogeneous gas: We start with the two-band model,
and in Sec. II(A) show how it relates to a one-band model with $p$-wave interactions.
In Sec. II(B), we describe the band structure and topology of the two-band model.
In Sec. III, we calculate the properties of trapped gases:
In Sec. III(A), we write the BdG equations and self-consistently calculate the order parameter and density.
In Sec. III(B), we visualize the MFs by calculating the DOS in real space and momentum space.
In Sec. III(C), we introduce MF operators and construct the localized MF states.
In Sec. III(D), we linearize the trap near the location of a MF,
finding an analytic expression for the localized MF wavefunction and
the gap between the MF state and other edge states. Finally we conclude in Sec. IV.

\section{homogeneous gas}
We start from the Hamiltonian of the 1D (pseudo) spin-$1/2$ Fermi gases
with chemical potential $\mu$,
\begin{eqnarray}
\label{ham}
H=\int\bigg(\Psi^\dagger(x)\big(H_0(x)-\mu\big)\Psi(x)\bigg)dx+H_I,
\end{eqnarray}
where $\Psi(x)=\big(\psi_\uparrow(x),\psi_\downarrow(x)\big)^\intercal$ annihilates
the spin-up and spin-down states. In an experiment, $\psi_\uparrow$
and $\psi_\downarrow$ correspond to two different hyperfine states of a
fermionic atom such as $^{40}K$. The single-particle Hamiltonian
$H_0(x)=-\frac{\hbar^2}{2m}\partial_x^2+\frac{i\hbar^2k_L}{m}\partial_x\sigma_z+
\frac{\hbar\Omega}{2}\sigma_x+E_r$ can be engineered by Raman lasers \cite{Spielman2011},
whose intensity is characterized by the Rabi frequency $\Omega$.
The recoil momentum of the Raman lasers is $\hbar k_L$,
$E_r=\frac{\hbar^2k_L^2}{2m}$ is the recoil energy,
and $\bm{\sigma}=(\sigma_x,\sigma_y,\sigma_z)$ is the vector of Pauli matrices.
For ultra-cold fermions, the interaction may be modeled by
$H_I=g_{1D}\int\psi_\uparrow^\dagger\psi_\downarrow^\dagger\psi_\downarrow\psi_\uparrow dx$, with coupling constant $g_{1D}$.
This coefficient can be related to the three-dimensional scattering length and the geometry of the
confinement \cite{Olshanii1998}. In a typical experiment, $|g_{1D}|\sim 70a_0E_r$ \cite{Hulet2010}, where $a_0$ is the Bohr radius.
We restrict ourselves to attractive interactions, $g_{1D}<0$.
We note that if we rotate our spin basis ($\sigma_x\rightarrow\sigma_z,\sigma_z\rightarrow\sigma_y$)
and identify $Z=\frac{\hbar\Omega}{2}$ as a Zeeman field,
and $\alpha=\frac{\hbar^2 k_L}{2m}$ as the SO coupling strength,
we recover the Hamiltonian of a semiconducting wire. Note $H_I$ is very different for a wire \cite{SSthoery}.
In the following sections we explore the physics of Eq. (\ref{ham}).
\subsection{One-band model}\label{oneband}
To get insight into Eq. (\ref{ham}), we first consider an approximation
where we truncate to a single band. We emphasize however that in all other sections, we work
with the full two-band Hamiltonian.

\begin{figure}[!htb]
\includegraphics[width=7cm]{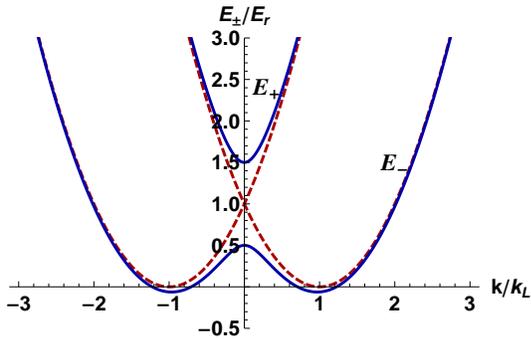}
\caption{(color online) Band structure of a 1D (pseudo) spin-$1/2$ gas.
The red (dashed) curves are the bare bands in the absence of SO coupling.
The blue (thick) curves are the upper band $E_+$ and
lower band $E_-$ in the presence of SO coupling, with the coupling strength $\hbar\Omega/E_r=1$.
\label{soband}}
\end{figure}

\begin{figure}[!htb]
\includegraphics[width=7.5cm]{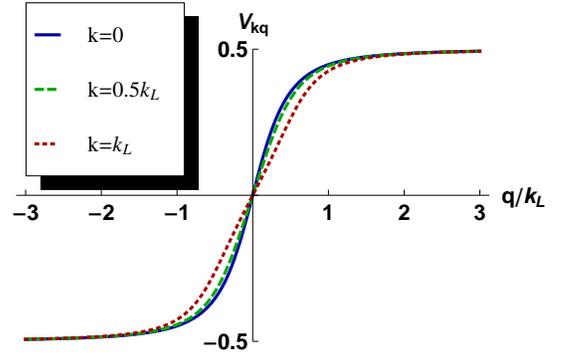}
\caption{(color online) Interaction coefficient $V_{kq}$
versus dimensionless momentum $q/k_L$ for $\hbar\Omega/E_r=2$.
The blue (thick), green (dashed) and red (dotted) curves correspond to $k=0,0.5k_L$ and $k_L$ respectively.
\label{interaction}}
\end{figure}

The physics of the single particle Hamiltonian is most transparent in momentum space,
$H=\sum_{k}\Psi^\dagger_k\left(\frac{\hbar^2k^2}{2m}+E_r-\frac{\hbar^2kk_L}{m}\sigma_z+
\frac{\hbar\Omega}{2}\sigma_x\right)\Psi_k$, where $\Psi_k=\left(\psi_{k\uparrow},\psi_{k\downarrow}\right)^\intercal$.
This Hamiltonian is readily diagonalized by
\begin{eqnarray}
\left(\begin{array}{c}
    d_k\\
    c_k\\
    \end{array}\right)
    &=&
    \left(\begin{array}{cc}
    {\rm cos}\frac{\theta_k}{2} & {\rm sin}\frac{\theta_k}{2}\\
    -{\rm sin}\frac{\theta_k}{2} & {\rm cos}\frac{\theta_k}{2}\\
    \end{array}\right)
    \left(\begin{array}{c}
    \psi_{k\uparrow}\\
    \psi_{k\downarrow}\\
    \end{array}\right)
\end{eqnarray}
with ${\rm tan}\theta_k=-\frac{m\Omega}{2k\hbar k_L}$, yielding
$H=\sum_k\left(E_-c_k^\dagger c_k+E_+d_k^\dagger d_k\right)$.
The energy spectrum has two helical bands, illustrated in Fig. \ref{soband}.
If the effective chemical potential $\tilde\mu=\mu-E_r\ll\frac{\hbar\Omega}{2}$,
only the lower band $E_-$ is filled with fermions.
Projecting the interactions into this band, we find
\begin{eqnarray}
\label{eint}
H_I^{1B}=\tilde g_{1D}\sum_{kqq'}\left(V_{kq}c^\dagger_{\frac{k}{2}+q}c^\dagger_{\frac{k}{2}-q}\right)
\left(V_{kq'}c_{\frac{k}{2}-q'}c_{\frac{k}{2}+q'}\right),
\end{eqnarray}
where $\tilde g_{1D}=g_{1D}/L_{1D}$, with $L_{1D}$ the length of the gas.
The fermionic anti-commutation relation, $c^\dagger_{\frac{k}{2}+q}c^\dagger_{\frac{k}{2}-q}=
-c^\dagger_{\frac{k}{2}-q}c^\dagger_{\frac{k}{2}+q}$,
implies that the interaction coefficient $V_{kq}$ is odd with respective to $q$,
$V_{kq}=\frac{1}{2}{\rm sin}\frac{\theta_{k/2+q}-\theta_{k/2-q}}{2}$.
At zero center of mass momentum, $V_q\equiv V_{k=0,q}=\frac{q}{2\sqrt{q^2+\hbar^2k_L^2\Omega^2/16E_r^2}}$.
In Fig. \ref{interaction}, we plot $V_{kq}$ as a function of $q$. The dependence on $k$
is weak for $k\lesssim k_L$.

The interaction in Eq. (\ref{eint}) is separable.
Given that $\tilde g_{1D}<0$, this interaction can lead to pairing with zero center of mass and an order parameter
$\Delta_q=\tilde g_{1D}V_q\sum_{q'}\langle V_{q'}c_{-q'}c_{q'}\rangle$,
where $\langle...\rangle\equiv\frac{{\rm Tr}(e^{-H/k_bT}...)}{{\rm Tr}(e^{-H/k_bT})}$ is the thermal average,
$k_b$ is the Boltzman constant and $T$ is the temperature.
The mean-field interaction becomes $H_I^{1B}=\sum_q\left(\Delta_qc_{q}^\dagger c^\dagger_{-q}
+\Delta_q^*c_{-q}c_q\right)-\tilde g_{1D}|\sum_qV_q\langle c_{-q}c_{q}\rangle|^2$.
By virtue of the symmetry of $V_q$,
the order parameter has a $p$-wave symmetry $\Delta_{-q}=-\Delta_q$. As is well established, such a $p$-wave
superfluid may possess Majorana edge modes \cite{Kitaev2001}. We will discuss these Majorana modes at length in
the two-band model.

\subsection{Two-band model}\label{twobands}
While the one-band model connects the SO coupled gases and $p$-wave superconductors,
we will focus on the richer two-band model in the remainder of the manuscript.
Within the mean-field approach, the interaction term is bilinear
\begin{eqnarray}
H_I&=&g_{1D}\int\psi_\uparrow^\dagger\psi_\downarrow^\dagger\psi_\downarrow\psi_\uparrow dx\\
&\approx&\int\left(\Delta(x)\left(\psi_\uparrow^\dagger\psi_\downarrow^\dagger+\psi_\downarrow\psi_\uparrow\right)
-\frac{\Delta(x)^2}{g_{1D}}\right)dx,
\end{eqnarray}
where the order parameter $\Delta(x)=g_{1D}\langle\psi_\downarrow\psi_\uparrow\rangle$ is assumed to be real.
Defining the operator $\tilde\Psi^\dagger(x)=\left(\psi_\uparrow^\dagger(x),
\psi_\downarrow^\dagger(x),\psi_\downarrow(x),\psi_\uparrow(x)\right)$,
the Hamiltonian can be written as,
\begin{eqnarray}
H=\int\left(\frac{1}{2}\tilde\Psi^\dagger(x)\mathcal{H}\tilde\Psi(x)-\frac{\Delta(x)^2}{g_{1D}}\right)dx
+\frac{1}{2}\left(T_-+T_+\right),
\end{eqnarray}
where
\begin{eqnarray}
\label{e8}
\notag\mathcal{H}&=&\left(-\frac{\hbar^2}{2m}\partial_x^2-\tilde\mu\right)\tau_z+\frac{i\hbar^2k_L}{m}\partial_x\tau_z\sigma_z\\
&+&\frac{\hbar\Omega}{2}\tau_z\sigma_x+\Delta(x)\tau_x\sigma_z,\\
T_{\pm}&=&{\rm Tr}\left(-\frac{\hbar^2}{2m}\partial_x^2-\tilde\mu\pm\frac{i\hbar^2k_L}{m}\partial_x\right).
\end{eqnarray}
The Pauli matrices $\bm{\sigma},\bm{\tau}$ operate in the spin subspace
and particle-hole subspace respectively,
\begin{eqnarray}
    \sigma_x&=&
    \left(\begin{array}{cccc}
      &  1  &  & \\
      1  &  &  & \\
      &  &  &  1\\
      &  &  1  & \\
    \end{array}\right),\tau_x=
    \left(\begin{array}{cccc}
      &  &  1  & \\
      &  &  &  1\\
      1  &  &  & \\
      &  1  &  & \\
    \end{array}\right)\\
    \sigma_z&=&
    \left(\begin{array}{cccc}
      1 &  &  & \\
      & -1 &  & \\
      &  & 1 & \\
      &  &  & -1\\
    \end{array}\right),\tau_z=
    \left(\begin{array}{cccc}
      1 &  &  & \\
      &  1 &  & \\
      &  & -1 & \\
      &  &  & -1 \\
    \end{array}\right).
\end{eqnarray}

\begin{figure}[!htb]
\includegraphics[width=4cm]{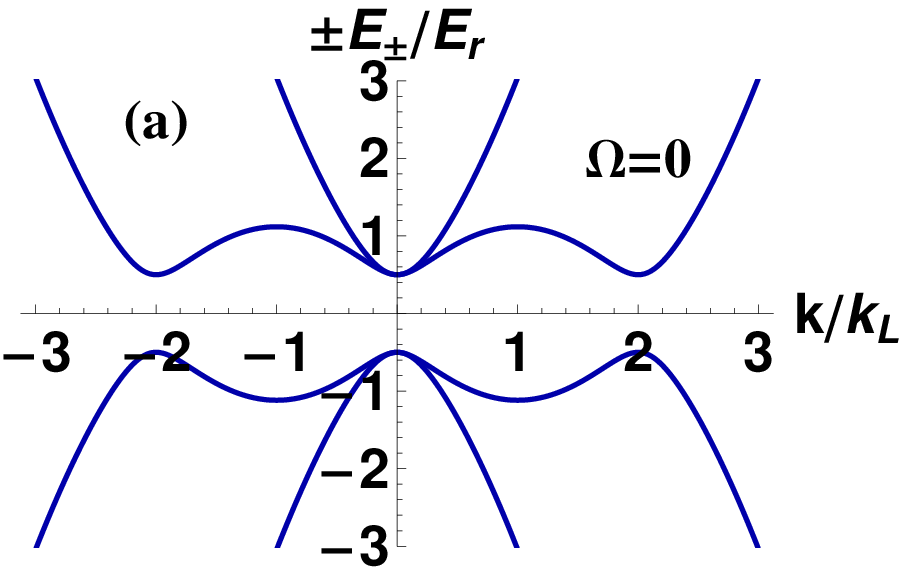}
\includegraphics[width=4cm]{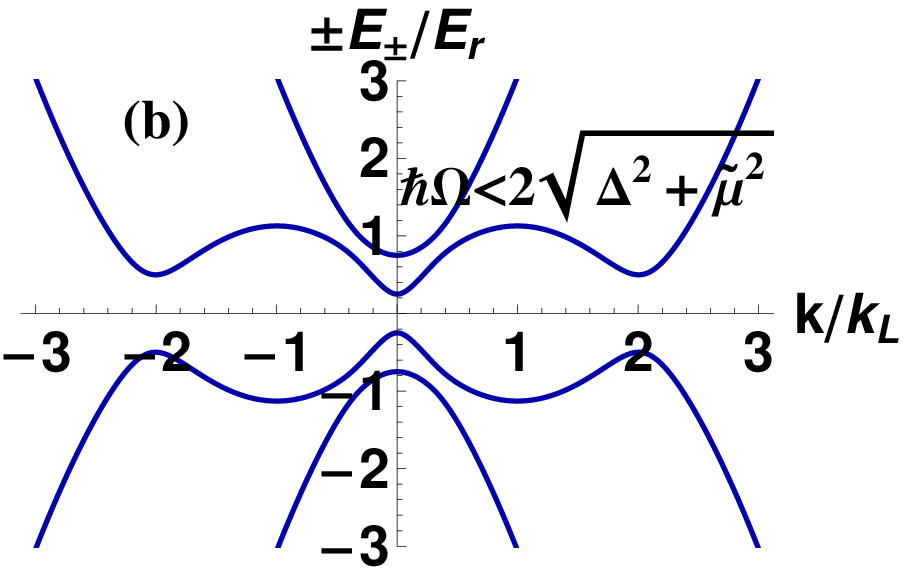}
\includegraphics[width=4cm]{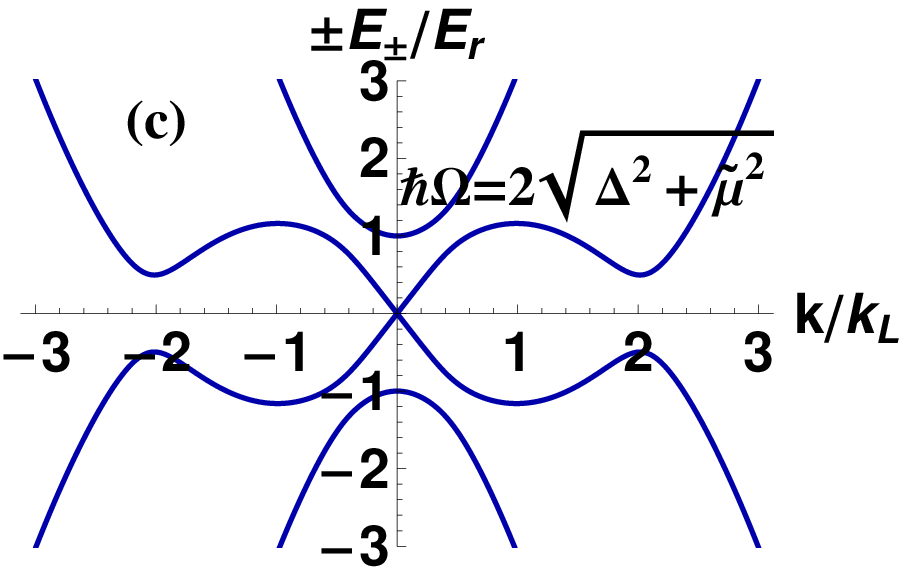}
\includegraphics[width=4cm]{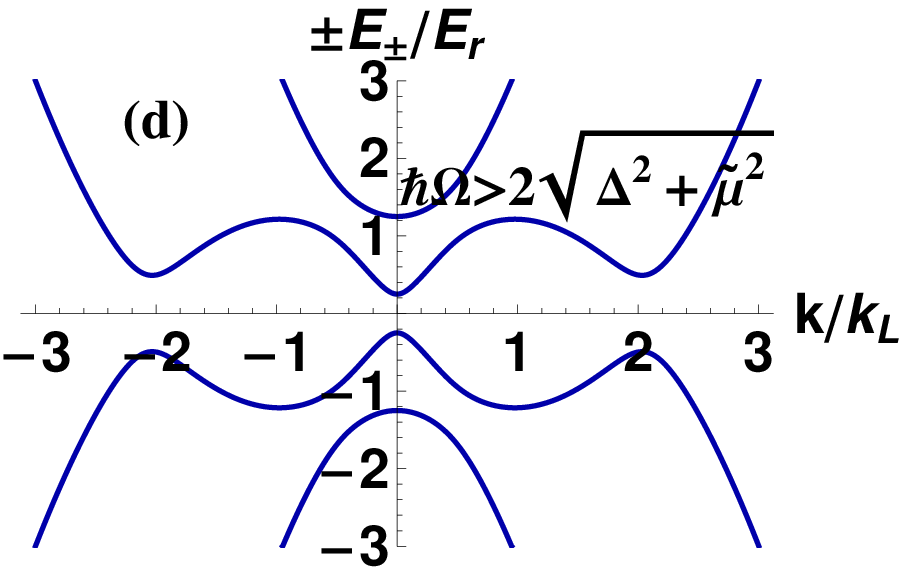}
\caption{(color online) Band structure of homogeneous gas.
From the top to the bottom, the four bands are $E_+,E_-,-E_-,-E_+$ respectively.
The parameters are $\mu=E_r,\Delta=0.5E_r$, and (a)$\Omega=0$, (b)$\Omega=0.5E_r$,
(c)$\Omega=E_r$, (d)$\Omega=1.5E_r$. \label{homobands}}
\end{figure}

The elementary excitations can be found by solving the BdG equations $\mathcal{H}W=EW$.
When $\Delta(x)=\Delta$ is spatially homogeneous, one can write the BdG equations in
momentum space as $\mathcal{H}_kW(k)=E(k)W(k)$,
where $\mathcal{H}_k$ is the $4\times4$ matrix produced by replacing $-i\partial_x\rightarrow k$ in Eq. (\ref{e8}).
The excitation spectrum $E(k)$ is most simply calculated by squaring $\mathcal{H}_k$ twice, and
extracting the characteristic polynomial \cite{Oreg2010}. This procedure yields
\begin{eqnarray}
\label{band}
\notag E_\pm^2(k)&=&\epsilon_0^2+2E_r\hbar^2k^2/m+\hbar^2\Omega^2/4+\Delta^2\\
&\pm&\hbar\sqrt{8E_r\epsilon_0^2k^2/m+\Omega^2\epsilon_0^2+\Omega^2\Delta^2},
\end{eqnarray}
where $\epsilon_0=\frac{\hbar^2k^2}{2m}-\tilde\mu$.
The four bands $E_+(k),E_-(k),-E_-(k),-E_+(k)$, as shown in Fig. \ref{homobands},
correspond to the four eigenvectors $W^{p+}(k)$, $W^{p-}(k)$, $W^{h-}(k)$, $W^{h+}(k)$.
The Hamiltonian $\mathcal{H}$ has the intrinsic symmetry, $\{\mathcal{H},\tau_y\}=1$.
Given two eigenvectors $W^{p\pm}$ with eigenvalues $E^\pm$,
one can always construct the other two $W^{h\pm}=i\tau_yW^{p\pm}$ with eigenvalues $-E^\pm$.
We therefore denote,
\begin{eqnarray}
\label{e3}
W^{p+}(k)&=&\bigl(u^+_{k\uparrow},u^+_{k\downarrow},-v^+_{k\downarrow},-v^+_{k\uparrow}\bigr)^\intercal\\
\label{e4}
W^{p-}(k)&=&\bigl(u^-_{k\uparrow},u^-_{k\downarrow},-v^-_{k\downarrow},-v^-_{k\uparrow}\bigr)^\intercal\\
\label{e5}
W^{h-}(k)&=&\bigl(v^-_{k\downarrow},v^-_{k\uparrow},u^-_{k\uparrow},u^-_{k\downarrow}\bigr)^\intercal\\
\label{e6}
W^{h+}(k)&=&\bigl(v^+_{k\downarrow},v^+_{k\uparrow},u^+_{k\uparrow},u^+_{k\downarrow}\bigr)^\intercal.
\end{eqnarray}
The unitary condition on the 4 by 4 matrix $\big(W^{p+}(k),W^{p-}(k),W^{h-}(k),W^{h+}(k)\big)$
also leads to the equalities $u^\pm_{k\downarrow}=(u^\pm_{-k\uparrow})^*$ and
$v^\pm_{k\downarrow}=-(v^\pm_{-k\uparrow})^*$.

In Fig. \ref{homobands}, the spectrum is shown for a range of parameters.
One important feature is the $k=0$ gap $E_0\equiv 2E_-(k=0)\equiv2|G|$,
where $G\equiv\sqrt{\tilde\mu^2+\Delta^2}-\hbar\Omega/2$. When $\Omega=0$, the two positive energy bands
touch at $k=0$, and $E_0>0$. The gas is in the
same universality class as a conventional $s$-wave superconductor.
Increasing the Raman laser strength such that $0<\hbar\Omega<2\sqrt{\Delta^2+\tilde\mu^2}$
separates the two bands and reduce $E_0$.
At $\hbar\Omega=2\sqrt{\Delta^2+\tilde\mu^2}$, $E_0$ is zero, and there is a topological transition.
Once $\hbar\Omega>2\sqrt{\Delta^2+\tilde\mu^2}$, $E_0$ is again positive,
but the gas is no longer a conventional superfluid, instead it has a non-trivial topological invariant.

The relevant topological invariant is a Berry phase.
Eqs. (\ref{e3}-\ref{e6}) can be thought of as maps from the real line ($-\infty\leq k\leq\infty$)
to the space of unit vectors in SU$(4)$. One can generate a closed path by taking
\begin{eqnarray}
\notag\mathcal{C}:&W^{p+}(-\infty)&\rightarrow W^{p+}(\infty)=W^{p-}(-\infty)\rightarrow\\
&W^{p-}(+\infty)&=W^{p+}(-\infty).
\end{eqnarray}
The equalities follow from noting that up to phases
\begin{eqnarray}
W^{p+}(-\infty)&=&W^{p-}(\infty)=\bigl(0,1,0,0\bigr)^\intercal\\
W^{p-}(-\infty)&=&W^{p+}(\infty)=\bigl(1,0,0,0\bigr)^\intercal.
\end{eqnarray}
Given this closed path, one can define the Berry phase
\begin{eqnarray}
\gamma&=&i\oint_\mathcal{C}\bm{W}^*\cdot\partial_k\bm{W}dk\\
\notag&=&i\int_{-\infty}^{\infty}\big(W^{p+}(k)\big)^*\cdot\partial_kW^{p+}(k)dk\\
&+&i\int_{-\infty}^{\infty}\big(W^{p-}(k)\big)^*\cdot\partial_kW^{p-}(k)dk.
\end{eqnarray}
In the case of a gauge which is not smooth, one would instead use
\begin{eqnarray}
\notag e^{i\gamma}&=&\lim_{\delta k\rightarrow0}\bigg(\prod_{k=-\infty}^{\infty}
\big(W^{p+}(k)\big)^*\cdot W^{p+}(k-\delta k)\\
&\times&\prod_{k=-\infty}^{\infty}\big(W^{p-}(k)\big)^*\cdot W^{p-}(k-\delta k)\bigg)\\
\notag&\times&\big(W^{p+}(-\infty)\big)^*\cdot W^{p-}(\infty)\big(W^{p-}(-\infty)\big)^*\cdot W^{p+}(\infty).
\end{eqnarray}
Since $H$ has real valued matrix elements, $e^{i\gamma}=\pm1$.
The Berry phase $\gamma$ is only well defined if the spectrum has no degeneracies.
We restrict $0\leq\gamma<2\pi$. For $G>0$, we find $\gamma=\pi$.
For $G<0$, $\gamma=0$. Somewhat counter-intuitively the $\gamma=0$ sector
corresponds to the ``topologically non-trivial"
state analogous to a 1D spinless $p$-wave superconductor.

\section{traps}
In this section we will solve the BdG equations for a trapped gas. The qualitative features of our results can be
anticipated by treating the system as locally homogeneous: the properties at position $x$ will be reminiscent of
those corresponding to a homogeneous gas with chemical potential $\tilde\mu(x)=\tilde\mu-V(x)$. Within this local
density approximation (LDA), one can define a function $G(x)=\sqrt{\tilde\mu(x)+\Delta(x)^2}-\hbar\Omega/2$,
where $G(x)=0$ corresponds to the boundaries between topologically distinct regions.
One expects there will be Majorana excitations at the boundaries.
We will use numerical solution of the BdG equation to explore this physics beyond the LDA. Further, in Sec. III(D) we will linearize
the BdG equations about the points $G(x)=0$, and analytically investigate these Majorana modes, without making a LDA.

\subsection{Order parameter and density}\label{density}
In the presence of a trap, we write the BdG equations in real space,
\begin{eqnarray}
\label{e1}
\mathcal{H}_{trap}W_n(x)&=&E_nW_n(x),
\end{eqnarray}
where
\begin{eqnarray}
\notag \mathcal{H}_{trap}&=&\left(-\frac{\hbar^2}{2m}\partial_x^2-\tilde\mu+V(x)\right)\tau_z\\
&+&\frac{i\hbar^2k_L}{m}\partial_x\tau_z\sigma_z+\frac{\hbar\Omega}{2}\tau_z\sigma_x+\Delta(x)\tau_x\sigma_z.
\end{eqnarray}
The eigenvectors $W_n(x)$ come in pairs, $W^{p}_n(x)$ and $W^{h}_n(x)$,
which correspond to eigenvalues $E_n\geq0$ and $-E_n$,
\begin{eqnarray}
W_n^{p}(x)&=&\bigl(u_{n\uparrow}(x),u_{n\downarrow}(x),v_{n\downarrow}(x),v_{n\uparrow}(x)\bigr)^\intercal\\
W_n^{h}(x)&=&\bigl(v^*_{n\uparrow}(x),v^*_{n\downarrow}(x),u^*_{n\downarrow}(x),u^*_{n\uparrow}(x)\bigr)^\intercal.
\end{eqnarray}
To make contact with our previous discussion, we note that
in the spatially homogeneous case, $n$ can be represented by a momentum $k_n$ and a sign $\varepsilon_n=\pm$,
so that $W_n^p(x)=e^{ik_nx}\big(u_{k_n\uparrow}^{\varepsilon_n},u_{k_n\downarrow}^{\varepsilon_n},
v_{k_n\downarrow}^{\varepsilon_n},v_{k_n\uparrow}^{\varepsilon_n}\big)^\intercal$ and
$W_n^h(x)=e^{ik_nx}\big((v_{k_n\uparrow}^{\varepsilon_n})^*,(v_{k_n\downarrow}^{\varepsilon_n})^*,
(u_{k_n\downarrow}^{\varepsilon_n})^*,(u_{k_n\uparrow}^{\varepsilon_n})^*\big)^\intercal$.
One then recovers Eqs. (\ref{e3})-(\ref{e6}).

Fixing $\{\tilde\mu,\Omega,g_{1D}\}$, we solve Eqs. (\ref{e1}) iteratively.
We discretize space into $n_{\rm grid}$ equally spaced points, and use a finite difference method
with a pseudo-spectral scheme
to represent $\mathcal{H}_{trap}$ as a $4n_{\rm grid}$ by $4n_{\rm grid}$ matrix.
In the $j$th iteration, we numerically diagonalize the matrix
$\mathcal{H}_{trap}^{(j)}$ with the order parameter $\Delta^{(j)}(x)$.
We start with a constant $\Delta^{(0)}(x)=\Delta_0$.
We extract the eigenvectors $W_n^{(j)}(x)$ and calculate the order parameter $\Delta^{(j+1)}(x)=
g_{1D}\sum_n\bigl(u_{n\downarrow}^{(j)}v^{*(j)}_{n\uparrow}\langle\xi_n\xi_n^\dagger\rangle
+v_{n\downarrow}^{*(j)}u_{n\uparrow}^{(j)}\langle\xi_n^\dagger\xi_n\rangle\bigr)$, where
$\xi_n$ is the annihilation operator of the Bogoliubov particle.
At temperature $T$, $\langle\xi_n^\dagger\xi_n\rangle=1/(e^{E_n/k_bT}+1)$.
Then we diagonalize $\mathcal{H}_{trap}^{(j+1)}$ and repeat the procedure.
We stop iterating when $|\Delta^{(j+1)}(x)-\Delta^{(j)}(x)|$ falls below a threshold.
The final convergent order parameter $\Delta^{(N)}(x)$ is largely independent of $n_{\rm grid}$
and $\Delta^{(0)}(x)$ when $n_{\rm grid}\geq1200$.
In the Appendix we explore the convergence with the real space grid size $n_{\rm grid}$.

\begin{figure}[!htb]
\includegraphics[width=8.5cm]{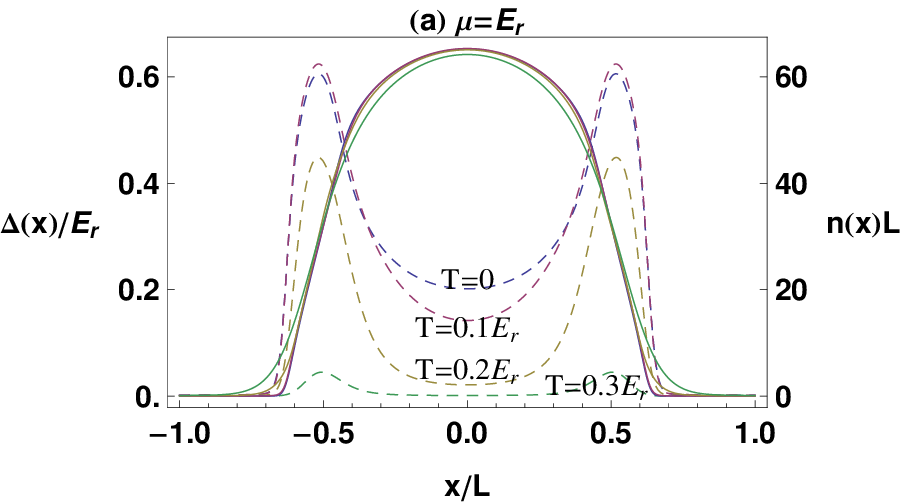}
\includegraphics[width=8.5cm]{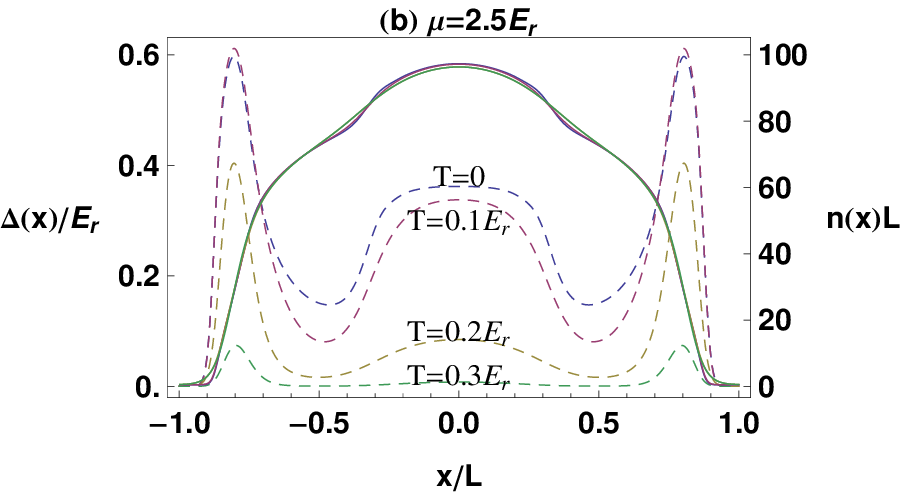}
\caption{(color online) Profiles of order parameter
$\Delta(x)=g_{1D}\sum_n\bigl(u_{n\downarrow}(x)v^*_{n\uparrow}(x)\langle\xi_n\xi_n^\dagger\rangle
+v_{n\downarrow}^*(x)u_{n\uparrow}(x)\langle\xi_n^\dagger\xi_n\rangle\bigr)$
(dashed curves) and density
$n(x)=\sum_{n\sigma}\bigl(|v_{n\sigma}(x)|^2\langle\xi_n\xi_n^\dagger\rangle+
|u_{n\sigma}(x)|^2\langle\xi_n^\dagger\xi_n\rangle\bigr)$
(solid curves) at temperatures $T=0,0.1E_r,0.2E_r,0.3E_r$.
Other parameters are $g_{1D}=-0.03E_rL,\hbar\Omega=2E_r,
\lambda=4,k_LL=100,(a)\mu=E_r,(b)\mu=2.5E_r$.
\label{profile}}
\end{figure}

The order parameters and density profiles for a gas in a harmonic trap $V(x)=\lambda(x/L)^2E_r$
are shown in Fig. \ref{profile}, where the dimensionless parameter $\lambda=4$ characterizes the stiffness of the trap,
and $2L$ is the simulation length with $k_LL=100$. We choose the
Rabi frequency to be $\hbar\Omega=2E_r$, and take $g_{1D}=-0.03E_rL$, corresponding to
the dimensionless interaction strength $\beta=m|g_{1D}|/\hbar^2n_0\sim2$, where $n_0$ is the central density at zero temperature.
For comparisons, experiments on 1D Fermi gases at Rice have $\beta\sim3$ \cite{Hulet2010}.
If $E_r/\hbar=50\rm{kHz}$ (a typical experimental value),
then these parameters correspond to a trap with small oscillation frequency $\omega=2\rm{kHz}$.
The order parameter has qualitatively different behavior if the center of the trap has
one or two bands occupied.
For relatively small chemical potential
$E_r-\sqrt{(\hbar\Omega/2)^2-\Delta(x)^2}\lesssim\mu\lesssim E_r+\sqrt{(\hbar\Omega/2)^2-\Delta(x)^2}$,
the center of the trap will be topologically non-trivial while the wings will be trivial.
This regime is illustrated in Fig. \ref{profile}(a).
The order parameter grows near the edge of the cloud. This is a feature of 1D where,
due to the divergence of the low energy density of state, the interactions are stronger for lower density \cite{Petrov2004}.
For $\mu\gtrsim E_r+\sqrt{(\hbar\Omega/2)^2-\Delta(x)^2}$, the center will be topologically trivial,
but there will be a band of the non-trivial state further out. Here the order parameter profile is quite
rich, with a central plateau, surrounded by two valleys and two peaks. The central plateau roughly corresponds to
where both bands are occupied. The order parameter is sensitive to temperature.
The bulk $\Delta$ is significantly suppressed and vanishes for $T\gtrsim0.2E_r$.
The density has no notable structure and is nearly independent of temperature for $T\lesssim 0.3E_r$.

\subsection{Density of states (DOS)}

\begin{figure}[!htb]
\includegraphics[width=8.5cm]{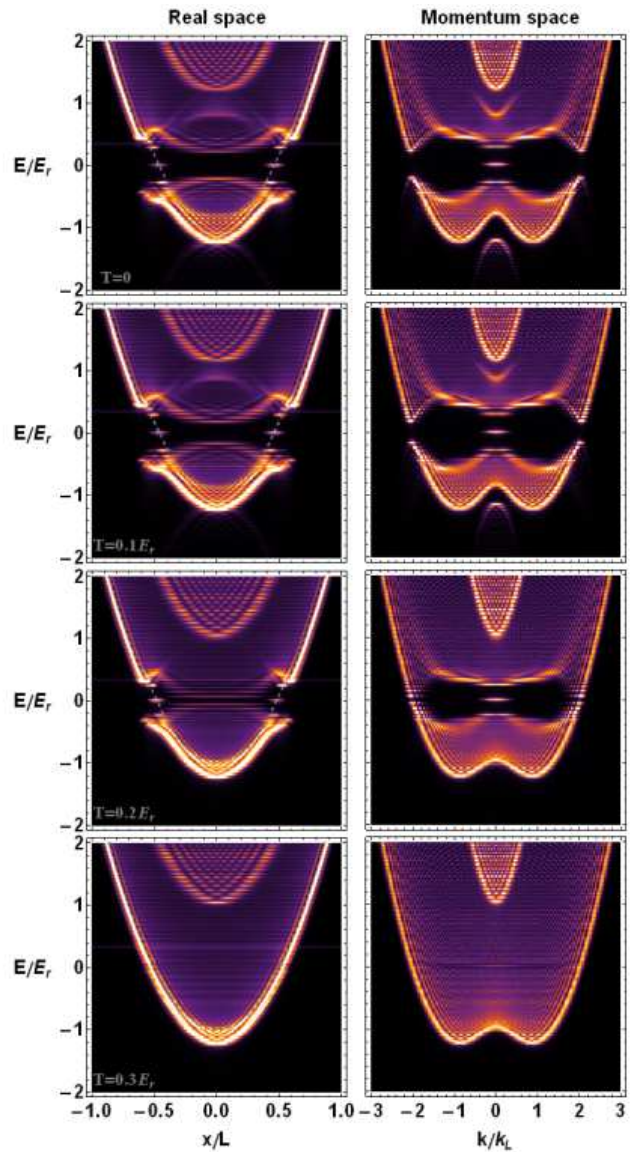}
\caption{(color online) Density of states (DOS) in real space (left panel)
and momentum space (right panel) at $T=0,0.1E_r,0.2E_r,0.3E_r$ from the top to the bottom,
with order parameters identical to those in Fig. \ref{profile}(a).
The grey (dashed) curves in the left panels is plotted with
$G(x)=\sqrt{\tilde\mu(x)^2+\Delta(x)^2}-\hbar\Omega/2$,
where the zero points of $G(x)$ pinpoint the position of MFs.
The brighter color corresponds to the higher spectral weight.\label{twomfs}}
\end{figure}

The elementary excitations are encoded in the single particle Green function
$G_{\sigma\sigma'}(x,t,x',t')=\frac{1}{i}\langle \hat T\psi_\sigma(x,t)\psi_{\sigma'}(x',t')\rangle$
and the associated spectral density $A_{\sigma\sigma'}(x,x',E)=2{\rm Im}\int e^{iEt}G_{\sigma\sigma'}(x,t,x',0)$,
where $\hat T$ is the time-ordering operator.
A local tunneling experiment can measure the density of states (DOS)
$\rho(E,x)=A_{\uparrow\uparrow}(x,x,E)+A_{\downarrow\downarrow}(x,x,E)$.
This quantity gives the number of single particle states with energy $E$ at position $x$.
It can be understood as an application of Fermi's Golden rule to the response to a tunneling probe. Within
our mean-field theory
\begin{eqnarray}
\rho(E,x)&=&\sum_{\sigma=\uparrow,\downarrow}\big(\rho^h_\sigma(-E,x)+\rho^p_\sigma(E,x)\big),
\end{eqnarray}
where
\begin{eqnarray}
\rho^h_\sigma(E,x)&=&\sum_n\big|v_{n\sigma}(x)\big|^2\delta(E_n-E)\\
\rho^p_\sigma(E,x)&=&\sum_n\big|u_{n\sigma}(x)\big|^2\delta(E_n-E).
\end{eqnarray}
We can similarly introduce a momentum resolved DOS
$\rho(E,k)=\int e^{ik(x-x')}\big(A_{\uparrow\uparrow}(x,x',E)+A_{\downarrow\downarrow}(x,x',E)\big)dxdx'$,
which can be measured with momentum resolved radio-frequency spectroscopy \cite{Zwierlein2012}. In the present case
\begin{eqnarray}
\rho(E,k)&=&\sum_{\sigma=\uparrow,\downarrow}\big(\rho_\sigma^h(-E,k)+\rho_\sigma^p(E,k)\big),
\end{eqnarray}
where
\begin{eqnarray}
\rho_\sigma^h(E,k)&=&\sum_n\bigg|\int v_{n\sigma}(x)e^{ikx}dx\bigg|^2\delta(E_n-E)\\
\rho_\sigma^p(E,x)&=&\sum_n\bigg|\int  u_{n\sigma}(x)e^{ikx}dx\bigg|^2\delta(E_n-E).
\end{eqnarray}

\begin{figure}[!htb]
\includegraphics[width=8.5cm]{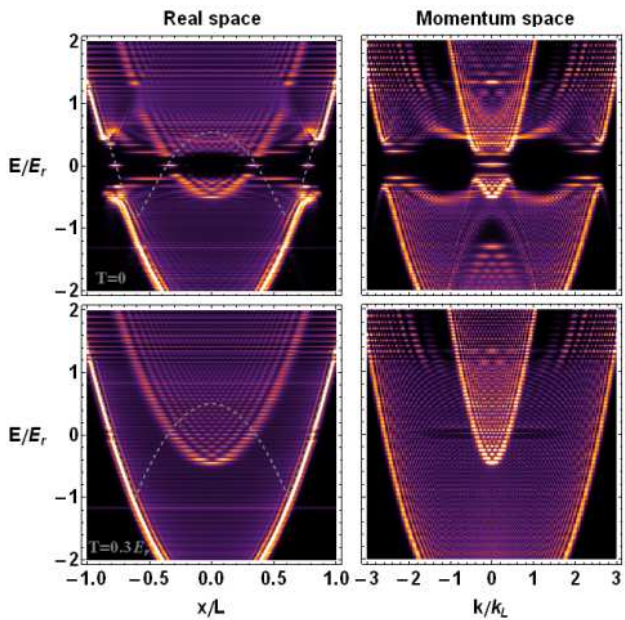}
\caption{(color online) Density of states (DOS) in real space (left panel)
and momentum space (right panel) at $T=0$ and $0.3E_r$,
with parameters identical to those in Fig. \ref{profile}(b).
The brighter color corresponds to the higher spectral weight.\label{fourmfs}}
\end{figure}

In Figs. \ref{twomfs} and \ref{fourmfs} we plot the DOS for the trapped gas with the
order parameters calculated in Sec. \ref{density}. We also show
a dashed curve corresponding to $G(x)=\sqrt{\tilde\mu(x)^2+\Delta(x)^2}-\hbar\Omega/2$.
The point where $G(x)=0$ represents the boundary between topologically distinct regions defined in Sec. \ref{twobands}.
For the parameters in Fig. \ref{twomfs}, $G(x)=0$ at two locations,
and we find that the BdG equations have two zero-energy modes, localized near these points. As
will be discussed later, these modes may be interpreted as MFs. They are clearly spectrally separated from all
other states. Fig. \ref{fourmfs} shows the case where $G(x)=0$ at four locations, representing four MFs.
The right panels of Figs. \ref{twomfs} and \ref{fourmfs} show the momentum space DOS.
The MF modes sit in a large gap at $k=0$.

As we have shown in Sec. \ref{density}, the order parameter decreases with temperature.
In real space, the bulk $\Delta$ becomes very small at $T=0.2E_r$, while $\Delta$ at the edges remains large:
the MFs at the edges are very clear for $T\lesssim0.2E_r$.
At $T=0.3E_r$, the order parameter is nearly zero and the gas becomes normal.

The evolution of the momentum space DOS parallels the real space DOS. As
temperature is increased from $T=0$, the gaps at large $k$ shrink.
The gap at $k=0$ remains robust until $T=0.3E_r$.

\begin{figure}[!htb]
\includegraphics[width=8.5cm]{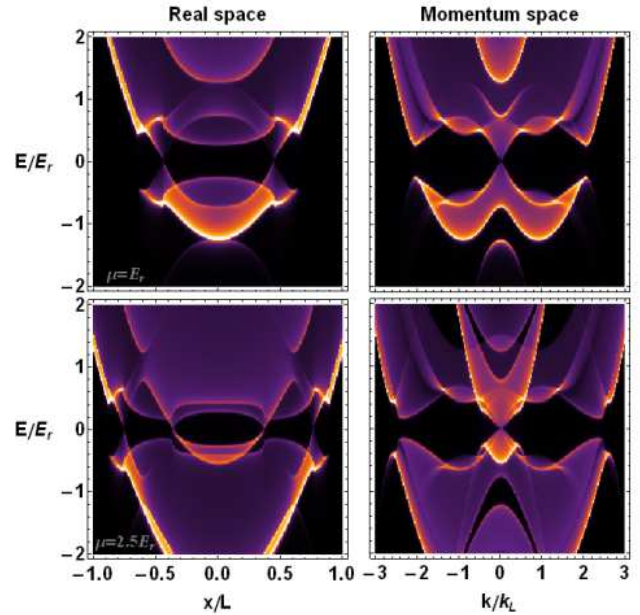}
\caption{(color online) Density of states (DOS) at zero temperature under the local density approximation (LDA).
The parameters are identical to those in Fig. \ref{profile}, except $\Delta(x)$ is calculated within the LDA.
The brighter color corresponds to the higher spectral weight.\label{lda}}
\end{figure}

Finally for comparison, we plot the DOS within a LDA. As illustrated in Fig. \ref{lda},
the LDA prediction for the DOS is qualitatively similar to the BdG result. The main
difference is that the LDA misses physics related to quantization. In particular, the zero energy modes
are not spectrally isolated in the LDA. They are, however, still located at roughly the same place in space.

\subsection{Majorana fermions (MFs)}\label{mf}
Here we explore the structure of the zero-energy states seen in Fig. \ref{twomfs}.
From our numerical solutions to the BdG equations, we have two wavefunctions
\begin{eqnarray}
\label{zeroenergy}
W_0^{p}(x)&=&e^{i\varphi_1'}\bigl(u_{0\uparrow}(x),u_{0\downarrow}(x),v_{0\downarrow}(x),v_{0\uparrow}(x)\bigr)^\intercal\\
W_0^{h}(x)&=&e^{i\varphi_2'}\bigl(v^*_{0\uparrow}(x),v^*_{0\downarrow}(x),u^*_{0\downarrow}(x),u^*_{0\uparrow}(x)\bigr)^\intercal,
\end{eqnarray}
which define operators
\begin{eqnarray}
\xi_0&=&\int\left(\big(W_0^p(x)\big)^\dagger\cdot\tilde\Psi(x)\right)dx\\
\xi_0^\dagger&=&e^{i(\varphi_2'-\varphi_1')}\int\left(\big(W_0^h(x)\big)^\dagger\cdot\tilde\Psi(x)\right)dx,
\end{eqnarray}
and obey $\mathcal{H}_{trap}W_0^p(x)\approx\mathcal{H}_{trap}W_0^h(x)\approx0$.
The phases $\varphi_1'$ and $\varphi_2'$ are not unique, and the factor in Eq. (\ref{zeroenergy})
must be introduced to make $\xi_0^\dagger$ conjugate to $\xi_0$. By construction
these are fermionic operators $\{\xi_0,\xi_0^\dagger\}=1$.

As zero-energy solutions to the BdG equations, both $\xi_0$ and $\xi_0^\dagger$ commute with $H$.
Hence the ground state is degenerate: $\xi_0|GS_1\rangle=0$ and $|GS_2\rangle=\xi_0^\dagger|GS_1\rangle$.
These two degenerate states can be used as a qubit for quantum information processing \cite{Sarma2008}.

The operator $\xi_0^\dagger$ which couples $|GS_1\rangle$ to $|GS_2\rangle$ is intrinsically nonlocal,
with weight at two spatially separated points. One can however introduce operators
\begin{eqnarray}
\chi_0=\frac{1}{\sqrt{2}}e^{i\varphi}\left(\xi_0+e^{-2i\varphi}\xi_0^\dagger\right)=\int f_0^\dagger(x)\cdot\tilde\Psi(x)dx\\
\bar\chi_0=\pm\frac{1}{\sqrt{2}i}e^{i\varphi}\left(\xi_0-e^{-2i\varphi}\xi_0^\dagger\right)=\int \bar f_0^\dagger(x)\cdot\tilde\Psi(x)dx
\end{eqnarray}
where
\begin{eqnarray}
f_0(x)&=&\frac{1}{\sqrt{2}}e^{i\varphi_1}\big(W_0^{p}(x)+e^{-i\varphi_2}W_0^{h}(x)\big)\\
\bar f_0(x)&=&\pm\frac{1}{\sqrt{2}i}e^{i\varphi_1}\big(W_0^{p}(x)-e^{-i\varphi_2}W_0^{h}(x)\big),
\end{eqnarray}
with arbitrary phases $\varphi_1=\varphi-\varphi_1'$ and $\varphi_2=2\varphi-\varphi_1'+\varphi_2'$.
By choosing the appropriate $\varphi_2$, these operators can be made local.
In particular if $G(x)=0$ at $x=x_1,x_2$, then $f_0(x)$ can be chosen to be nonzero only near $x_1$, and
$\bar f_0(x)$ only near $x_2$.

The operators $\chi_0$ and $\bar \chi_0$ obey the Majorana algebra: $\chi_0^\dagger=\chi_0$,
$\bar\chi_0^\dagger=\bar\chi_0$, $\{\chi_0,\bar\chi_0\}=0$, $\{\chi_0,\chi_0\}=\{\bar\chi_0,\bar\chi_0\}=1$.
They commute with the Hamiltonian.

Note, as we will use in the next subsection,
$f_0(x)\equiv e^{i\varphi_f}\big(u_{f\uparrow}(x),u_{f\downarrow}(x),v_{f\downarrow}(x),v_{f\uparrow}(x)\big)$
obeys the BdG equations, but the resulting Bogoliubov transformation is not unitary as it
changes the commutation relations. Since $\chi_0=\chi_0^\dagger$, we have $u_{f\sigma}(x)=e^{-2i\varphi_f}v_{f\sigma}^*(x)$.
For smaller systems, coupling between these modes push them away from $E_0=0$.

\subsection{Eigen-energies of excited states near a MF}\label{linearize}
As seen in Figs. \ref{twomfs}-\ref{fourmfs}, the MFs are localized in real space and momentum space.
Thus we can calculate their properties by linearizing the trap around
their locations in position space, and linearizing momentum around $k=0$.
As previously discussed, the locations of the MFs can be found via the LDA. There are generally four MFs, localized at
$x_m=\pm L\sqrt{R_{\pm m}/\lambda E_r}$,
where $R_{\pm m}\equiv\tilde\mu\pm\sqrt{\hbar^2\Omega^2/4-\Delta_m^2}$,
with $\Delta_m\equiv\Delta(x=x_m)$. We restrict ourselves to the location of one MF, $x_m=L\sqrt{R_{+m}/\lambda E_r}$.
We write the linearized BdG Hamiltonian as the sum of two terms
$\mathcal{H}_{lin}=\mathcal{H}_0+\mathcal{H}_i$,
\begin{eqnarray}
\mathcal{H}_0&=&\frac{\hbar\Omega}{2}\tau_z\sigma_x+\Delta_m\tau_x+\sqrt{\hbar^2\Omega^2/4-\Delta_m^2}\tau_x\sigma_z\\
\mathcal{H}_i&=&\tilde\lambda(x-x_m)\tau_z-\kappa\tau_z\sigma_z,
\end{eqnarray}
where $\tilde\lambda=2\lambda x_mE_r/L^2$ and $\kappa=\hbar^2k_Lk/m$.
The ``interaction" term $\mathcal{H}_i$ can be treated as a perturbation,
and it vanishes as $x\rightarrow x_m,k\rightarrow0$.
In the absence of perturbations, $\mathcal{H}_{lin}=\mathcal{H}_0$ has two
degenerate zero-energy states
\begin{eqnarray}
\mathcal{D}_1&=&\frac{\sqrt{2}}{2}({\rm sin}\phi,-{\rm cos}\phi,-{\rm cos}\phi,{\rm sin}\phi)\\
\mathcal{D}_2&=&\frac{\sqrt{2}}{2}(-{\rm cos}\phi,{\rm sin}\phi,-{\rm sin}\phi,{\rm cos}\phi),
\end{eqnarray}
where ${\rm sin}\phi=\sqrt{(\hbar\Omega/2+\Delta_m)/\hbar\Omega}$.
Following the standard approach to first-order degenerate perturbation theory,
we diagonalize the Hamiltonian projected into
the subspace $\{\mathcal{D}_1,\mathcal{D}_2\}$,
\begin{eqnarray}
\label{e10}
\mathcal{\bar H}_{lin}=\left(\begin{array}{c}
 \mathcal{D}_1\\
 \mathcal{D}_2
\end{array}
\right)H_{lin}\big(\mathcal{D}_1,\mathcal{D}_2\big)=K\bar\sigma_z+X\bar\sigma_x,
\end{eqnarray}
where $K=-2\hbar\Delta_mk_Lk/m\Omega$ and $X=-4\lambda E_rx_m(x-x_m)R_{+m}/\hbar\Omega L^2$.
The Pauli matrices $\bar{\bm\sigma}$ operate in the subspace $\{\mathcal{D}_1,\mathcal{D}_2\}$.
Noting that $\big[X,K\big]=iC$ with $C=16\sqrt{\lambda}E_r^{3/2}R_{+m}^{3/2}\Delta_m/\hbar^2\Omega^2k_LL$,
one can define the operators $a=\frac{K-iX}{\sqrt{2C}},a^\dagger=\frac{K+iX}{\sqrt{2C}}$
that satisfy $\big[a,a^\dagger\big]=1$. The eigen-equations of $\mathcal{\bar H}_{lin}$ then become
\begin{eqnarray}
\label{e7}
\frac{\sqrt{2C}}{2}\left(
\begin{array}{cc}
 -(a^\dagger+a) & i(a^\dagger-a) \\
 i(a^\dagger-a) & a^\dagger+a
\end{array}
\right)
\left(\begin{array}{c}
 \bar u_n \\
 \bar v_n
\end{array}\right)
 =\bar E_n\left(\begin{array}{c}
 \bar u_n \\
 \bar v_n
\end{array}
\right)
\end{eqnarray}
where $\bar u_n=\mathcal{D}_1\cdot W_n,\bar v_n=\mathcal{D}_2\cdot W_n$.
Combining $\bar u_n,\bar v_n$ gives the equations
\begin{eqnarray}
\label{e9}
-\sqrt{2C}\left(
\begin{array}{cc}
 0 & a^\dagger \\
 a & 0
\end{array}
\right)
\left(\begin{array}{c}
\bar u_n+i\bar v_n \\
\bar u_n-i\bar v_n
\end{array}
\right)=
\bar E_n\left(\begin{array}{c}
\bar u_n+i\bar v_n \\
\bar u_n-i\bar v_n
\end{array}
\right).
\end{eqnarray}
Squaring Eq. (\ref{e9}) yields harmonic oscillator Hamiltonian, and allows one to read off
\begin{eqnarray}
\label{e2}
\bar E_n=\pm\sqrt{2C}\sqrt{n}\quad(n=0,1,2,...).
\end{eqnarray}
Not only is there a zero-energy mode $\bar E_0=0$ (the Majorana mode), but there is a ladder of localized
fermionic modes, whose energy spacing is proportional to $\lambda^{1/4}$, and whose wavefunction
components are excited harmonic oscillator states.
For a homogeneous gas where $\lambda=0$, the energy spacing becomes zero.
\begin{figure}[!htb]
\includegraphics[width=7.5cm]{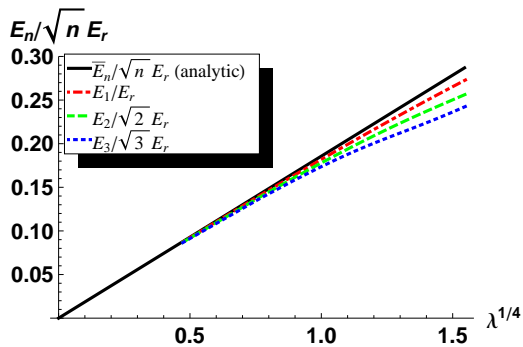}
\caption{(color online) The gap between the MF state and excited states as a function of trap stiffness
$\lambda^{1/4}$: the trapping potential is $V(x)=\lambda(x/L)^2E_r$.
The black (thick) curve is plotted based on the analytic Eq. (\ref{e2}).
The red (dot-dashed), green (dashed), blue (dotted) curves are the energy levels
$E_1/E_r,E_2/\sqrt{2}E_r,E_3/\sqrt{3}E_r$ respectively.
They are numerically calculated from Eq. (\ref{e1}) with the parameters identical to the thick curve.
\label{analytical}}
\end{figure}

In Fig. \ref{analytical}, we plot $\bar E_n/\sqrt{n}E_r$ as a function of $\lambda^{1/4}$ (black thick curve)
based on Eq. (\ref{e2}), and compare to the numerical results calculated from Eq. (\ref{e1}).
The dot-dashed (red), dashed (green), and dotted (blue) curves show the energy levels of the first three
excited states. We see the analytic results agree well with the numerics for small $\lambda$.
For larger $\lambda$, the corrections to Eq. (\ref{e10}) are important,
and the discrepancy between the analytic and numerical results becomes notable, especially for larger $n$.

At $n=0$ ($\bar E_0=0$), the zero-energy mode has wavefunction
\begin{eqnarray}
\bar u(x)&=&\frac{1}{2\sigma\sqrt{\pi}}e^{-(x-x_m)^2/2\sigma^2}\\
\bar v(x)&=&\frac{1}{2i\sigma\sqrt{\pi}}e^{-(x-x_m)^2/2\sigma^2},
\end{eqnarray}
where the width $\sigma=\sqrt{\Delta_mLE_r^{1/2}/R_{+m}^{3/2}k_L\lambda^{1/2}}$ is proportional to $\lambda^{-1/4}$.

\section{Summary}
We have investigated a (pseudo) spin-$1/2$ spin-orbit (SO) coupled Fermi gas in a one-dimensional geometry.
We first relate this system to a one-band model with $p$-wave interactions.
We then described the band structure and calculated the Berry phase $\gamma$ of the full two-band model.
We found $\gamma$ distinguishes two topologically distinct sectors, with $\gamma=\pi$
corresponding to a conventional superconductor.
By self-consistently solving the Bogoliubov-de Gennes equations and calculating
both the position resolved and momentum resolved density of states,
we visualized the Majorana fermion (MF) states in real and momentum space at finite temperatures.
These spectra can be probed using the position resolved or momentum resolved radio-frequency spectroscopy
\cite{Ketterle2007,Zwierlein2012}. We introduced MF operators and constructed the localized MF states.
We further linearized the trap near the location of a MF,
finding an analytic expression for the localized MF wavefunction and
the gap between the MF state and other edge states.

This physics can be experimentally studied in a bundle of weakly coupled tubes containing fermionic atoms \cite{Hulet2010}.
By applying appropriate Raman lasers to these quasi-1D tubes \cite{Zhang2012,Zwierlein2012},
one can produce an array of quasi-1D SO coupled Fermi clouds.
Our calculations show that the MFs can be observed in such settings.

There are, however, significant experimental challenges. Most notably, the Raman induced
SO coupling relies on the ability of optical photons to flip the atomic hyperfine spin.
As Spielman argues \cite{Spielman2009}, if the Raman lasers are detuned by a frequency $\Delta$ from
an excited state multiplet (and $\hbar\Delta$ is large compared to the fine structure splitting $A_f$), then
the coupling strength $\Omega$ scales as $1/\Delta^2$.
(This is contrasted with typical AC stark shifts, which instead scale as $1/\Delta$.
The extra suppression is due to quantum interference between the amplitudes arising from different intermediate states.)
The rate of inelastic light scattering $\Gamma_i$ also scales as $1/\Delta^2$. The ratio $\upsilon=\Gamma_i/\Omega$
is therefore roughly independent of detuning. In terms of microscopic parameters, $\upsilon\propto\hbar/A_f\tau$,
where $\tau$ is the lifetime of the excited states. For $^6Li$, $\hbar/A_f\tau\sim5.8\times10^{-4}$, for $^{40}K$,
$\hbar/A_f\tau\sim3.5\times10^{-6}$ and for $^{87}Rb$, $\hbar/A_f\tau\sim8.3\times10^{-7}$.
One sees $^{40}K$ has a much longer lifetime than $^{6}Li$ in a SO coupled Fermi experiment.
The situation is even less favorable at the typical magnetic field $\sim830G$ \cite{Ketterle2005}
where one encounters Feshbach resonances in $^{6}Li$. The large magnetic field
decouples the electron spin and the nuclear spin, and the relevant hyperfine states
effectively only differ by their nuclear spin. As a result, the Raman laser couplings vanish between these states.
However for $^{40}K$, the typical resonance field $\sim200G$ \cite{Jin2004} is much smaller,
and the relevant hyperfine states have larger Raman couplings.
We therefore expect $^{40}K$ is a promising candidate for producing an interacting SO coupled Fermi gas.

\section{Acknowledgement}
The work in Sec. \ref{oneband} is largely derived from notes by Bhuvanesh Sundar.
The authors thank Randall Hulet for discussions of experimental difficulties.
R. W. is supported by CSC, the NNSFC, the NNSFC
of Anhui (under Grant No. 090416224), the CAS, and the
National Fundamental Research Program (under Grant
No. 2011CB921304). This material is based upon work supported by the
National Science Foundation under Grant No. PHY-1068165 and a grant from the Army
Research Office with funding from the DARPA OLE program.

\section{Appendix}
In this Appendix, we explore the convergence of our self-consistent calculations
with the grid spacing $\delta x=2L/n_{\rm grid}$.
We show how the energy and the order parameter
for a zero-temperature homogeneous gas in
a box of size $2L$ with periodic boundary conditions depends on $n_{\rm grid}$.

Within our mean-field theory, the energy of this homogeneous gas is
\begin{eqnarray}
\label{geq}
E_g=\sum_k\bigg(\epsilon_0(k)-\frac{1}{2}\big(E_+(k)+E_-(k)\big)\bigg)-\frac{|\Delta|^2}{\tilde g_{1D}},
\end{eqnarray}
where $\tilde g_{1D}=g_{1D}/2L,\epsilon_0(k)=\frac{\hbar^2k^2}{2m}-\tilde\mu$,
and $E_\pm(k)$ is the excitation spectrum given in Eq. (\ref{band}).
The summation index $k$ is discretized as
$k=-\frac{n_{\rm grid}}{2L}\pi,-\left(\frac{n_{\rm grid}-2}{2L}\right)\pi,
...,\left(\frac{n_{\rm grid}-4}{2L}\right)\pi,\left(\frac{n_{\rm grid}-2}{2L}\right)\pi$,
and Eq. (\ref{geq}) can be calculated numerically.

\begin{figure}[!htb]
\includegraphics[width=7.5cm]{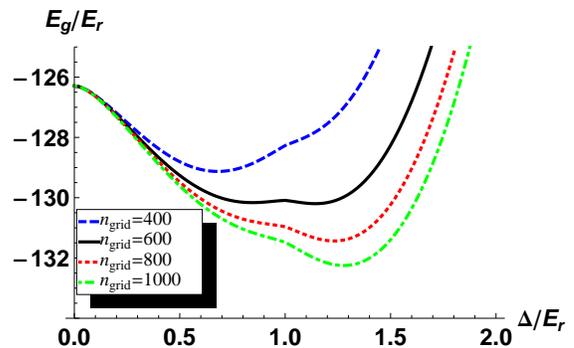}
\caption{(color online) Ground state energy $E_g/E_r$ versus order parameter $\Delta/E_r$.
The blue (dashed), black (thick), red (dotted), green (dot-dashed) curves
correspond to $n_{\rm grid}=400,600,800,1000$ respectively.
Other parameters are $\tilde g_{1D}=-0.02E_r,\hbar\Omega=2E_r,\lambda=0,k_LL=100,\mu=E_r$.
\label{groundstate}}
\end{figure}

Fig. \ref{groundstate} shows $E_g$ as a function of $\Delta$ for
$n_{\rm grid}=400,600,800,1000$.
We find non-trivial behavior at intermediate $n_{\rm grid}$. In particular,
for these parameters and $n_{\rm grid}=600$, the energy has two local minima,
and the gap equations has four solutions, corresponding $\Delta=0$ and other three stationary points.
Such behavior is an artifact of the discretization, as it goes away for $n_{\rm grid}\gtrsim800$.
It does, however, indicate that in the presence of an appropriate tuned optical lattice,
there will be metastable superfluid states.

\begin{figure}[!htb]
\includegraphics[width=7.5cm]{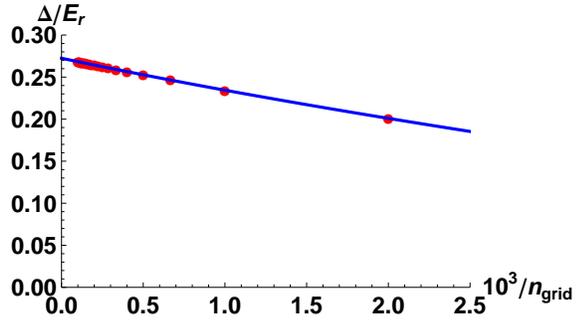}
\caption{(color online) Order parameter $\Delta/E_r$ versus $10^3/n_{\rm grid}$.
The red dots are calculated from Eq. (\ref{minimize}). The blue (thick) curve is
an extrapolation. The parameters here are identical to those in Fig. \ref{profile}(a) except for $\lambda=0$.
\label{grid}}
\end{figure}

In Fig. \ref{grid}, we show how the order parameter $\Delta$ depends on $n_{\rm grid}$.
We calculate $\Delta$ by minimizing $E_g$,
\begin{eqnarray}
\label{minimize}
\frac{\partial E_g}{\partial|\Delta|}\bigg|_{|\Delta|>0}=0.
\end{eqnarray}
We see $\Delta$ converges to a finite value as $n_{\rm grid}\rightarrow\infty$.
For the simulation size $n_{\rm grid}=1200$ used in the main text, the finite grid error is
$\frac{|\Delta(n_{\rm grid}=\infty)-\Delta(n_{\rm grid}=1200)|}{\Delta(n_{\rm grid}=\infty)}\leq12\%$.

\end{document}